\documentclass[fleqn,twoside]{article}
\usepackage{espcrc2,epsf,epsfig,graphicx}

\newcommand{\AmS}{{\protect\the\textfont2
  A\kern-.1667em\lower.5ex\hbox{M}\kern-.125emS}}

\title{High Energy  Astrophysical Tau  Neutrinos: 
       The Expectations\thanks{Talk given at 12th International Symposium on Very
        High Energy Cosmic Ray Interactions (XII ISVHECRI), 15$-$20 July,
        2002, Geneva, Switzerland.}}

\author{H. Athar
        \\Physics Division,
        \\ National Center for Theoretical Sciences,
 101 Section 2, Kuang Fu Road, Hsinchu 300, Taiwan}

\begin{document}

\begin{abstract}
    Aspects related to production, propagation and prospects for
    observations
    of high energy astrophysical tau neutrinos originating from some representative
    extra terrestrial sources
    such as atmosphere of earth, our galactic plane as well as possibly from distant
    sites of gamma ray bursts in the energy range
    $10^{3} \leq E/\mbox{GeV} \leq 10^{11}$ are reviewed.

\vspace{1pc}
\end{abstract}

\maketitle

\section{INTRODUCTION}

A search for high energy astrophysical tau neutrinos will not only
yield quite useful information about some of the highest energy
phenomenons occurring  in the present universe related to the
origin of observed high energy gamma rays and  ultra high energy
cosmic rays but also can possibly probe physics beyond standard
model of particle physics.

    The representative extra terrestrial sources of high energy astrophysical tau
    neutrinos with energy greater than $ 10^{3}$ GeV include the
    atmosphere of earth, the plane of our galaxy, the center of distant
    (active) galaxies as well as possibly the sites of gamma ray bursts (GRBs).
     Here, I
    summarize the present status of intrinsic flux estimates for high energy
    astrophysical tau neutrinos, both originating from the source itself as well as
    during propagation (of ultra high energy cosmic rays, assumed to be dominantly
    protons here). In light of recently growing empirical evidence for
    neutrino flavor oscillations, I will mention the effects of
    neutrino flavor mixing on this intrinsic flux (in two flavor approximation,
    for simplicity). The presently
    envisaged prospects for possible observations of  intrinsic and oscillated high energy
    astrophysical tau neutrino flux through several different detection strategies
    will also be
    presented. For a recent brief review on high energy astrophysical
    neutrinos, see \cite{Athar:2002im}.

\section{HIGH ENERGY ASTROPHYSICAL TAU NEUTRINOS}
\subsection{Production}

    High energy astrophysical tau neutrino production can occur in
    $pp$ and $p\gamma $ interactions
    taking place in cosmos. These interactions produce unstable hadrons such as
    $D_{S}$ and  $B_{S}$. At further higher center of mass energy, production of
     other heavier states such as $t\bar{t}$, $W^{*}$ as well as
     $Z^{*}$ is also possible.  All of these states decay into
     $\nu_{\tau}$ directly or indirectly, with $D_{S}$ being the lightest one to decay
     directly into $\nu_{\tau}$. Copious production of $\pi^{\pm}$ also occurs in
     these interactions, which decay into non tau neutrinos.
       The absolute normalization of
     high energy
     astrophysical tau neutrinos is fixed in a similar way as for
     high energy astrophysical muon neutrinos. In general,
     the $D_{S}$ dominates the tau neutrino production over
     the entire energy range under consideration here.
     In these interactions,
     the target protons and
     photons are considered to be  present inside the source as well as in the interstellar
     medium between the source and the detector. The
     high energy tau neutrinos possibly originating from GRBs
     are an example of the former
     situation, whereas the galactic plane and GZK tau neutrinos are examples
     for the latter situation.
     The GZK tau neutrinos are a result of neutrino flavor
     oscillations of GZK muon neutrinos, which are produced in
     $p\gamma \to \Delta \to \pi^{\pm}X$ interactions \cite{Engel:2001hd}.
        The atmospheric tau neutrino flux is estimated in
        \cite{Pasquali:1998xf}, whereas the galactic tau neutrino flux is estimated
         and compared with atmospheric one in
        \cite{Athar:2001jw}. In these estimates, the tau neutrino flux is
        obtained by solving  the system of coupled cascade equations that
        describes
        the propagation of protons, unstable hadrons and leptons in the
        presence of  a varying  target density  medium.

    Depending upon the distance to the source and the concerned energy range,
    the production of
    high energy tau neutrinos may or may not become comparable to the high energy
    muon neutrinos. For instance, in case of galactic plane, it is
    the intrinsic muon neutrino flux that dominates, however, it
    is not the case for atmosphere of earth for the whole energy range
    under consideration here. In the atmosphere of
    earth, for $E > 10^{5}$ GeV, intrinsic  tau neutrino production is quite
    comparable to intrinsic muon neutrino production as both originate from $D$'s.

\subsection{Oscillations during propagation}

The empirical evidence for neutrino flavor mixing is now rather
compelling.  In particular, the explanation of recent
statistically significant data concerning the atmospheric muon
neutrino deficit is suggestive of $\nu_{\mu} \to \nu_{\tau}$
flavor oscillations. In the context of two neutrino flavors, the
oscillation probability formula is
\begin{equation}
\label{one}
    P(\nu_{\mu}\rightarrow \nu_{\tau})=\sin^{2} 2\theta
    \cdot \sin^{2}(l/l_{\rm osc}).
\end{equation}
Here $\theta $ is neutrino flavor mixing angle between $\nu_{\mu}$
and $\nu_{\tau}$, whereas $l_{\rm osc}\equiv \pi E/\delta m^{2}$.
Presently, the global best fit values of $\sin^{2}2\theta $ and $\delta
m^{2}$ are  essentially $\sim 1$  and $\sim 2.5\cdot 10^{-3}$
eV$^{2}$, respectively \cite{Valle:2002tm}. In Eq. (\ref{one}), $l$ is the
distance between the source and the detector.  In case of complete
averaging, namely when $l \gg l_{\rm osc}$, the second $\sin^{2} $
factor is
    equal to 0.5. Taking $\sin^{2}2\theta \sim 1$, I obtain, $P \leq 0.5$.
     This is substantial, as compared to the intrinsic
    production probability of high energy astrophysical tau neutrinos, which
    is rather suppressed relative to that for muon
    neutrinos in some distant astrophysical sites as mentioned in the last section.
    High energy astrophysical tau
neutrinos thus seem to be roughly as abundant as high energy
muon neutrinos except from the atmosphere of earth, given the
present status of {\em neutrino flavor mixing}. For a description of
three (and four) neutrino flavor mixing effects for high energy
astrophysical tau neutrinos, see \cite{Athar:2000yw}.

Equation 1 ignores the effect of a possible enhancement in $\theta
$ due to (coherent forward) neutrino scattering  during 
propagation.
\begin{figure}[htb]
\includegraphics[width=0.455\textwidth]{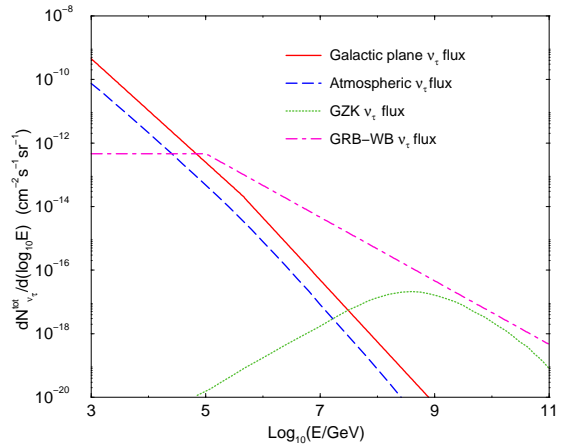}
\caption{Galactic plane, horizontal atmospheric and GZK high
energy tau neutrino flux under the assumption of (two) neutrino
flavor mixing. For comparison, tau neutrino flux in a fireball model
 of GRB 
is also shown \cite{Waxman:1997ti}. The energy range shown covers
all the presently envisaged high energy (tau) neutrino detectors.}
\label{fig:l}
\end{figure}
This approximation is presently justified, given the current
status of high energy astrophysical tau neutrino observations
\cite{Athar:1999gw}. Further discussion on propagation effects of
mixed high energy (tau) neutrinos can be found in
\cite{Lunardini:2000sw}, whereas for unmixed high energy tau
neutrinos propagation effects (mainly related to observations),
see \cite{Naumov:1998sf}. Given the intrinsic neutrino fluxes,
${\rm d}N^{{\rm int}}_{\nu_{\mu}}/{\rm d(log_{10}}E)$ and ${\rm
d}N^{\rm int}_{\nu_{\tau}}/{\rm d(log_{10}}E)$ at the production site,
the total tau neutrino flux ${\rm d}N^{\rm tot}_{\nu_{\tau}}/{\rm
d(log_{10}}E)$ arriving at the detector, including the effects of
neutrino flavor oscillations during the propagation, is
\begin{eqnarray}
  {\rm d}N^{\rm tot}_{\nu_{\tau}}/{\rm d(log_{10}}E)  =
  P\cdot {\rm d}N^{\rm int}_{\nu_{\mu}}/{\rm d(log_{10}}E) \nonumber \\
   +(1-P)\cdot {\rm d}N^{\rm int}_{\nu_{\tau}}/{\rm d(log_{10}}E),
\end{eqnarray}
where $P\equiv P(\nu_{\mu}\to \nu_{\tau})$ is given by Eq.
\ref{one}.  Figure 1 displays the ${\rm d}N^{\rm
tot}_{\nu_{\tau}}/{\rm d(log_{10}}E)$ from some representative
extra terrestrial sources.

Let me emphasize here a simple point that neutrino flavor mixing
is an intrinsic property of the neutrino state. It has nothing to
do with the energy of the neutrino. It is the probability of
neutrino oscillation that depends on neutrino energy.

\subsection{Prospects for possible observations}

    Present status of currently operating and under planning
high energy (tau) neutrino detectors is described in
\cite{Resvanis}. In general, the high energy astrophysical tau
neutrino flux arrives at an earth based detector in three general
directions.

\subsubsection{Downward going}

    It might be possible to search for downward going high energy
    tau neutrinos through {\em double shower} technique
    \cite{Learned:1994wg}. Quantitative details of this suggestion
    for a typical Km$^{3}$ instrumented volume (ice) Cherenkov
    radiation detector were given in \cite{Athar:2000rx}.
    The event simulations done by IceCube collaboration based on
    \cite{Athar:2000rx} can be found in \cite{icecube}. In principle, this
    technique may also be feasibly implemented in other alternative 
 detection  methods as well.
     A high energy tau neutrino flux at the level of $\sim 10^{-11}
     ({\rm cm}^{2}\cdot {\rm s}\cdot {\rm sr})^{-1}$ gives
several double shower type events per year irrespective of its
origin in an instrumented volume of Km$^{3}$ for an  incident
energy of several $10^{6}$ GeV.
\begin{figure}[htb]
\includegraphics[width=0.50\textwidth]{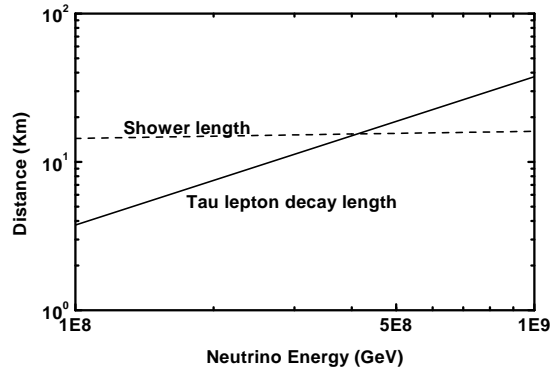}
\caption{Comparison of the tau lepton decay length and the shower
length of the first shower (defined as twice the depth at maximum)
in air.} \label{fig:2}
\end{figure}
\begin{figure}[htb]
\includegraphics[width=.5\textwidth]{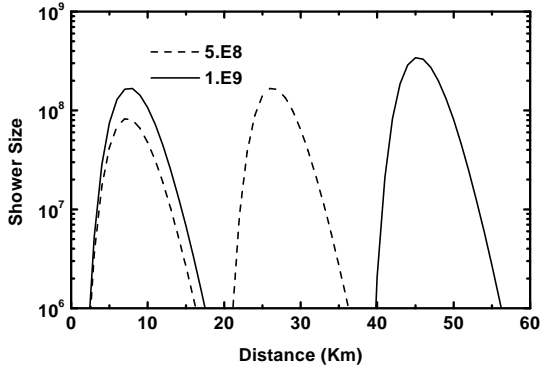}
\caption{Typical longitudinal development of a double shower
produced by the high energy tau neutrino nucleon charged current
deep inelastic scattering occurring inside Pierre Auger array.}
\label{fig:3}
\end{figure}

    For the energy range between $10^{3}$ GeV to $5 \cdot 10^{5}$ GeV, the two
    showers will be quite difficult to separate from each other
    for typical Cherenkov radiation detectors. 
 Between $5\cdot 10^{5}$ GeV and $10^{6}$ GeV,
    the proposed Megaton detectors are a possibility, whereas between
     $2\cdot 10^{7}$ GeV and $5 \cdot 10^{8}$ GeV, detectors with instrumented volume
     of more than Km$^{3}$ may be needed. On the other hand for energy
    greater than $10^{9}$ GeV, the two showers are too distant from
    each other (essentially hundreds of Km), so are difficult to
    track for presently planned (single) earth based detectors
    \cite{Fargion:1997eg}.

\subsubsection{Quasi horizontal}

    The possibility of observation of near horizontal
    high energy tau neutrinos through double shower technique
    for Pierre Auger array
    was briefly considered in \cite{Athar:2000tg}.
    Following \cite{Athar:2000rx}, an approximately half an order of
    magnitude energy interval, namely
    $5\cdot 10^{8}\leq E/{\rm GeV}\leq 10^{9}$,  was obtained by
    requiring that the longitudinal profile of the two showers do
    not overlap in the array length, taken to be $\sim $ 60 Km
    (see Fig. \ref{fig:2} and Fig. \ref{fig:3} for illustration).
        The two showers are considered to develop in the air.
    For energy above $10^{9}$ GeV, the decay length of the tau lepton in air
    is larger than
    the size of the array and the double shower signature can not be observed.
    However, the energy at which the two showers start separating
    in Pierre Auger array lies below the presently planned threshold for
    high energy neutrino detection.
    The crucial factor here being the {\em difference} in the
    incident tau neutrino energy dependence on the spread and
    separation of the two showers. The double shower type
    event rate for Pierre Auger array can be estimated
    following \cite{Athar:2000rx}.

\subsubsection{Upward going}

    The upward going tau neutrino flux is suppressed by the deep
    inelastic tau neutrino nucleon scattering inside the earth for $E\geq 5\cdot
    10^{4}$ GeV \cite{Halzen:1998be}. In fact, for upward going
    high energy tau neutrinos which cross almost all the diameter of
    the earth, the suppression in the incident flux is essentially
    exponential.

    A variant of the last two situations is a scenario in which
     the high energy tau neutrino nucleon charged current deep inelastic
     interaction occurs
    just once inside the earth and a tau lepton is produced. This
    tau  lepton after exiting the earth
     produces an air shower which
    might possibly be measured in a future large earth, plane,
    balloon or space based detector \cite{Domokos:1997ve}.

\section{CONCLUSIONS}
$\bullet$
    So far, there are only few explicit estimates for intrinsic high energy
    astrophysical tau neutrino flux. Existing ones are for
    atmosphere of earth and for our galaxy.

$\bullet$    Neutrino flavor oscillations is an interesting
possibility to
    expect $\nu_{\tau}$  from $\nu_{\mu}$. For
    distant astrophysical tau neutrino sources in the energy range
    $10^{3} \leq E/\mbox{GeV} \leq 10^{11}$, presently
$
    {\rm d}N^{\rm tot}_{\nu_{\tau}}/{\rm d(log_{10}}E)  \leq 0.5\cdot
    {\rm d}N^{\rm int}_{\nu_{\mu}}/{\rm d(log_{10}}E).
$

$\bullet$
    High energy astrophysical tau neutrino observation can possibly be
    achieved in future large arrays through several different
    detection strategies.

 The author thanks Physics Division of National Center for
        Theoretical Sciences for  support.

\end{document}